\documentclass[preprint,prb,aps,superscriptaddress]{revtex4}
\usepackage{graphicx}
\usepackage{color}
\usepackage{amssymb}
\usepackage{units}
\usepackage[latin1]{inputenc} 

\graphicspath{{fig/}}
\begin{document}

\title
{Magnetic anisotropies and magnetization reversal of the
Co$_2$Cr$_{0.6}$Fe$_{0.4}$Al Heusler compound}

\author{J. Hamrle}
\affiliation{Fachbereich Physik and Forschungsschwerpunkt MINAS,
Technische Universit\"at Kaiserslautern,
Erwin-Schr\"odinger-Stra\ss e 56, D-67663 Kaiserslautern, Germany}

\author{S. Blomeier}
\affiliation{Fachbereich Physik and Forschungsschwerpunkt MINAS,
Technische Universit\"at Kaiserslautern,
Erwin-Schr\"odinger-Stra\ss e 56, D-67663 Kaiserslautern, Germany}

\author{O. Gaier}
\affiliation{Fachbereich Physik and Forschungsschwerpunkt MINAS,
Technische Universit\"at Kaiserslautern,
Erwin-Schr\"odinger-Stra\ss e 56, D-67663 Kaiserslautern, Germany}

\author{B. Hillebrands}
\affiliation{Fachbereich Physik and Forschungsschwerpunkt MINAS,
Technische Universit\"at Kaiserslautern,
Erwin-Schr\"odinger-Stra\ss e 56, D-67663 Kaiserslautern, Germany}

\author{R. Sch\"afer}
\affiliation{IFW Dresden, Helmholtzstra\ss e 20, D-01069 Dresden,
Germany}

\author{M. Jourdan}
\affiliation{Institut f\"ur Physik,
Johannes-Gutenberg-Universit\"at, Staudinger Weg 7, D-55128,
Mainz, Germany}

\date{June 08, 2006}
\begin{abstract}
Magnetic anisotropies and magnetization reversal properties of the
epitaxial Heusler compound Co$_2$Cr$_{0.6}$Fe$_{0.4}$Al (CCFA)
deposited on Fe and Cr buffer layers are studied. Both samples
exhibit a growth-induced fourfold anisotropy, and magnetization
reversal occurs through the formation of stripy domains or
90$^\circ$ domains. During rotational magnetometric scans the
sample deposited on Cr exhibits about 2$^\circ$ sharp peaks in the
angular dependence of the coercive field, which are oriented along
the hard axis directions. These peaks are a consequence of the
specific domain structure appearing in this particular measurement
geometry. A corresponding feature in the sample deposited on Fe is
not observed.
\end{abstract}

\maketitle

\section{Introduction}

Studies of ferromagnetic (FM) half-metals are mainly driven by
their possible applications for spintronic devices as a potential
source of a 100\% polarized spin current. Some Heusler alloys are
promising candidates due to their high Curie temperature and
expected half-metallicity even for partially disordered systems.

Among the many Heusler systems already studied, the compound
Co$_2$Cr$_{0.6}$Fe$_{0.4}$Al (CCFA) attracted significant
experimental \cite{elm03,aut03,fel03,kel05,hir05,con06,blo06} and
theoretical \cite{miu04,ant05,wur06} attention. CCFA is an
interesting candidate due to its high Curie temperature of
\unit[760]{K} \cite{fel03} and its high value of volume
magnetization of $\approx$\unit[3]{$\mu_B$} per formula unit
\cite{con06,ino06} at \unit[5]{K} (the theoretical value is
\unit[3.8]{$\mu_B$} per formula unit \cite{gal02}). At room
temperature, CCFA exhibits a magnetoresistance of 30\% for the
pure Heusler compound in powder form \cite{blo03} or 88\% if
artificial Al$_2$O$_3$ grain boundaries \cite{blo06} are used.
Simple Co$_2$Cr$_{0.4}$Fe$_{0.6}$Al/Cu/Co$_{90}$Fe$_{10}$
trilayers showed a large giant magnetoresistance of 6.8\% at RT,
\cite{kel05} while a tunneling magnetoresistance ratio of 52\% at
RT and 83\% at 5K was reported for the system
CCFA/AlO$_x$/Co$_{75}$Fe$_{25}$. \cite{ino06}

In this article we study magnetic anisotropies and magnetization
reversal properties of epitaxial CCFA films deposited on Fe and Cr
buffer layers.

\section{Sample properties and preparation}

We have investigated two samples consisting of CCFA sputtered onto
an epitaxial Fe or Cr buffer layer:
Al(\unit[2.5]{nm})/CCFA(\unit[80]{nm})/Cr(\unit[8]{nm})/MgO(001)
(in the following called CCFA/Cr) and
Al(\unit[4]{nm})/CCFA(\unit[105]{nm})/Fe(\unit[10]{nm})/MgO(001)
(in the following called CCFA/Fe). The buffer layers were
deposited by electron beam evaporation onto a single-crystalline
MgO(001) substrate, while the epitaxial CCFA films were
subsequently deposited by dc magnetron sputtering. A more detailed
description of the sample preparation procedure can be found
elsewhere. \cite{con06} The films grow with the B2 structure, as
there is full disorder between the Cr and Al positions, but order
on the Co positions. \cite{con06} A volume magnetization of
$\mu_\mathrm{CCFA}\approx2.5\mu_B$ per formula unit was measured
at \unit[300]{K}. \cite{con06} 4-circle XRD scans yielded axes
lengths of \unit[$a=0.569\pm0.003$]{nm},
\unit[$b=c=0.578\pm0.006$]{nm} (\unit[$a=0.570\pm0.005$]{nm},
\unit[$b = c = 0.583\pm0.012$]{nm}) for CCFA/Fe (CCFA/Cr), where
the $a$-axis is perpendicular to the sample surface and $b$ and
$c$ are in-plane axes. Therefore, in-plane strain is as small as
$\approx0.8\pm2.6$\%.

\section{Magnetic anisotropies}

\begin{figure}
\begin{center}
\includegraphics[width=0.6\textwidth]{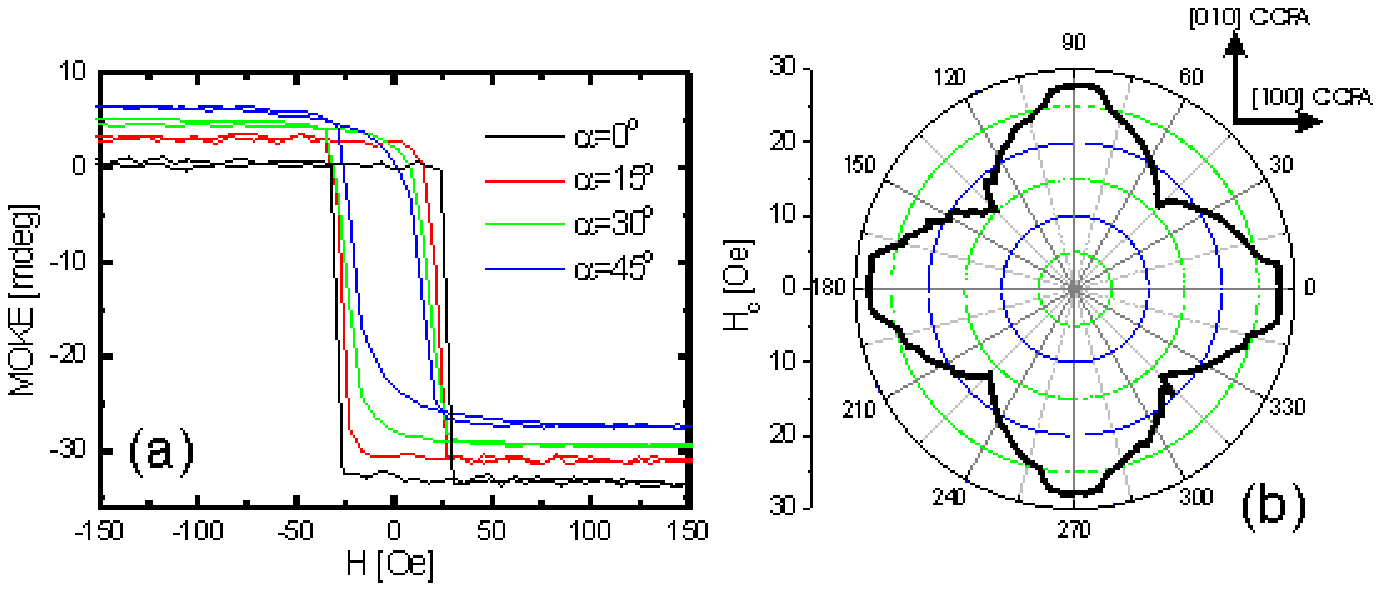}
\end{center}
\caption{%
\label{f:fe:loops}%
(a) (color on-line) MOKE loops of CCFA/Fe for different sample
orientations. (b) polar plot of the coercivity $H_c$.}
\end{figure}

\begin{figure}
\begin{center}
\includegraphics[width=0.6\textwidth]{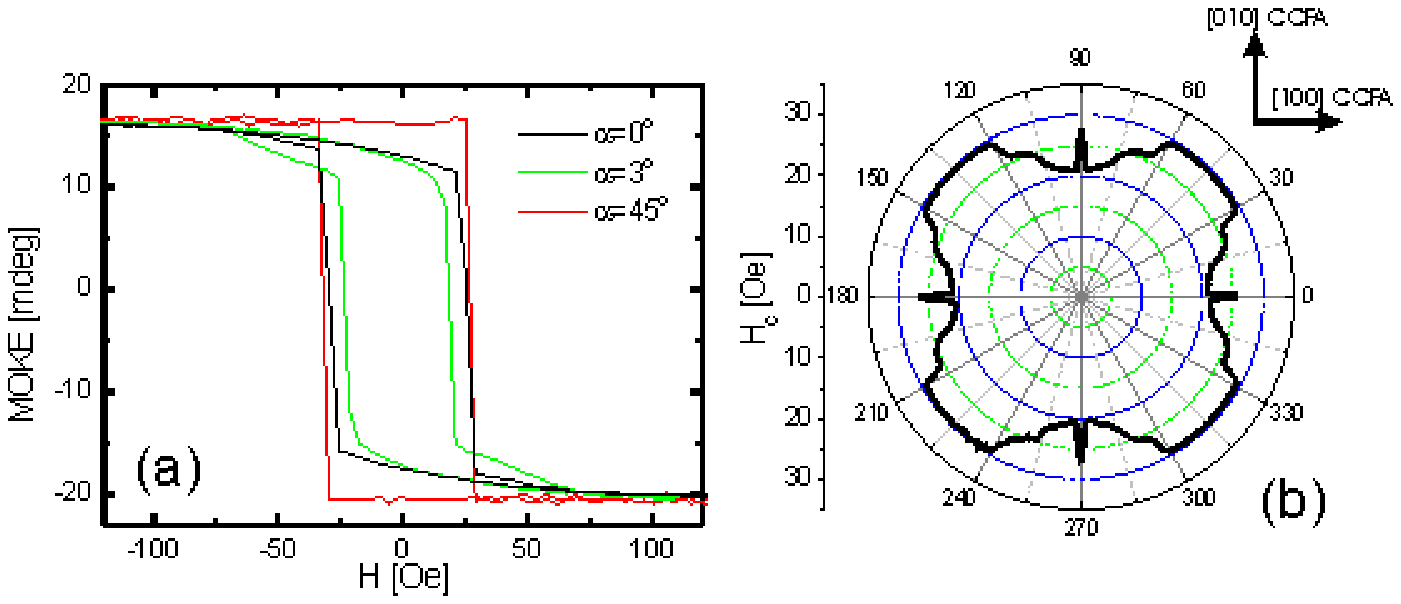}
\end{center}
\caption{%
\label{f:cr:loops}%
(a) (color on-line) MOKE loops of CCFA/Cr for different sample
orientations. (b) polar plot of the coercivity $H_c$.}
\end{figure}

Magnetic anisotropies were investigated using magneto-optical Kerr
effect (MOKE) magnetometry. All measurements were performed in
longitudinal geometry with an angle of incidence near 45$^\circ$.
The MOKE hysteresis loops were measured as a function of the
in-plane angle $\alpha$ of the applied magnetic field with respect
to the in-plane [100] direction of the CCFA film by rotating the
sample about its normal axis and recording the loop every
1$^\circ$. Typical loops as well as the corresponding polar plots
of the obtained coercivity for CCFA/Cr and CCFA/Fe are presented
in Figs.~\ref{f:fe:loops} and \ref{f:cr:loops}. In both figures,
the [100] CCFA axis (i.e. the [110] axis of the MgO
substrate\cite{con06}) corresponds to a sample orientation of
$\alpha = 0^\circ$. As it can be seen in the figure, both the
CCFA/Cr and CCFA/Fe samples show a fourfold in-plane anisotropy
reflecting the crystallographic symmetry of the CCFA film. Several
interesting features can be observed:

(1) The maximum coercivities for CCFA/Fe and CCFA/Cr are 28 and
\unit[29]{Oe}, respectively, and are thus very similar.

(2) The in-plane easy axes lie along the $\langle$100$\rangle$
CCFA axes for CCFA/Fe whereas in case of CCFA/Cr they are rotated
by 45$^\circ$, i.e. oriented along the $\langle$110$\rangle$ CCFA
axes. At the moment, the reasons for this behavior are not clear.
According to previous studies, both types of film grow with the
same crystallographic orientation and thus should exhibit the same
orientation of their magnetic easy axes. \cite{mat95}

(3) CCFA/Cr exhibits unique sharp peaks in $H_c$ as a function of
the in-plane angle $\alpha$ as narrow as 2$^\circ$. These peaks
are aligned parallel to the $\langle$100$\rangle$ CCFA hard axes.
As it will be shown in the next section, they originate from a
peculiar magnetization reversal mechanism when $H$ is applied
exactly parallel to the $\langle$100$\rangle$ CCFA hard axis.
Similar peaks have been observed on epitaxial bcc Fe on Cu(001):
\cite{scheurer93} in this case the origin of the peaks was
attributed to the presence of structural microdomains in Fe,
having different easy and hard axes and being magnetically coupled
to each other. However, as it will be discussed later, in our case
of CCFA/Cr the peaks originate from magnetic frustration effects
during magnetization reversal.

\section{Magnetization reversal}

Magnetization reversal mechanisms were studied using a Kerr
microscopy setup described in Ref. \cite{hubert}. Using different
orientations of a nearly crossed polarizer and analyzer we are
sensitive to magnetization components both parallel and transverse
to the externally applied field $H$. When the orientations of the
analyzer and the polarizer are almost equal to 0$^\circ$ and
90$^\circ$, the Kerr microscope is sensitive to the longitudinal
magnetization component (i.e. the component parallel to the
incidence plane of light). On the other hand, when the
orientations of the analyzer and the polarizer are nearly equal to
45$^\circ$ and -45$^\circ$, the microscope is sensitive to the
transverse magnetization component (i.e. the component transverse
to the plane of incidence).

\begin{figure}
\begin{center}
\includegraphics[scale=1]{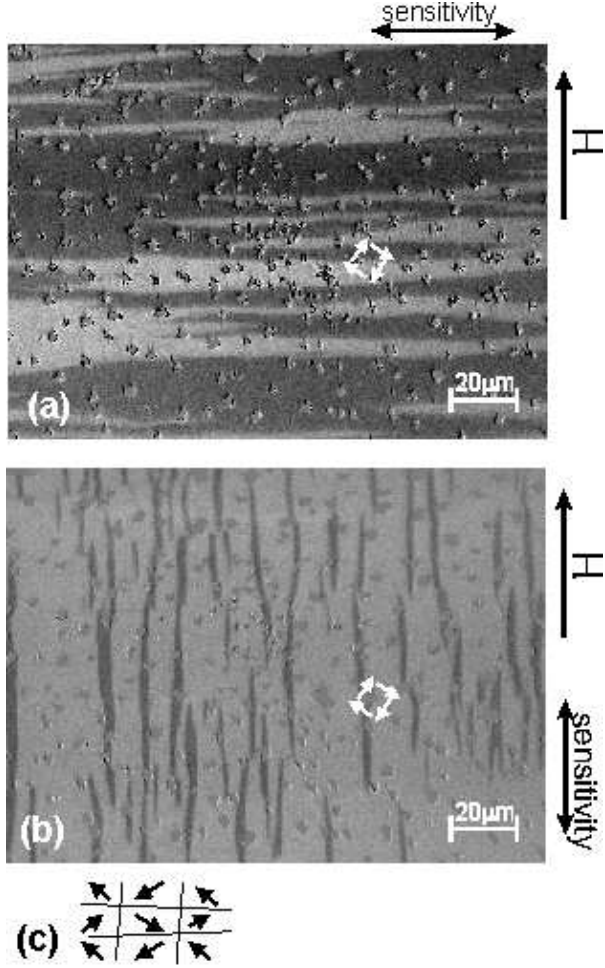}
\end{center}
\caption{%
\label{f:cr:demag}%
Kerr microscopy image of the demagnetized state of CCFA/Cr. The
demagnetizing AC field was applied in hard axis direction
($\alpha=0^\circ$). The image corresponds to sensitivity to
magnetization components (a) transverse and (b) parallel to $H$,
respectively. Both images show an identical domain pattern,
recorded under different sensitivity conditions. Magnetization
directions are indicated by small white arrows. (c) Sketch of the
underlying 90$^\circ$ domain structure.}
\end{figure}

The demagnetized state of CCFA/Cr is displayed in
Fig.~\ref{f:cr:demag}. The images shown in panels (a) and (b) are
Kerr images of an identical domain structure, but with sensitivity
to the magnetization components transverse and parallel to $H$,
respectively. The sample was demagnetized by an AC field
(frequency \unit[$\approx$10]{Hz}, maximal amplitude
\unit[$\approx80$]{Oe}) applied in hard axis direction
($\alpha=0^\circ$) whose amplitude was gradually decreased to
zero. Both (a) and (b) Kerr images show a stripy domain pattern,
with the stripe directions being parallel to the respective
direction of the measured magnetization components. Therefore, the
demagnetized state exhibits a 90$^\circ$ domain structure, as it
is sketched in panel (c). The dots in the Kerr images are small
particles on the sample surface, which do not influence domain
propagation or nucleation. The demagnetized state of CCFA/Fe is
very similar to that of CCFA/Cr (it also exhibits a 90$^\circ$
domain structure), so we do not discuss it here any further.

\begin{figure}
\begin{center}
\includegraphics[scale=1]{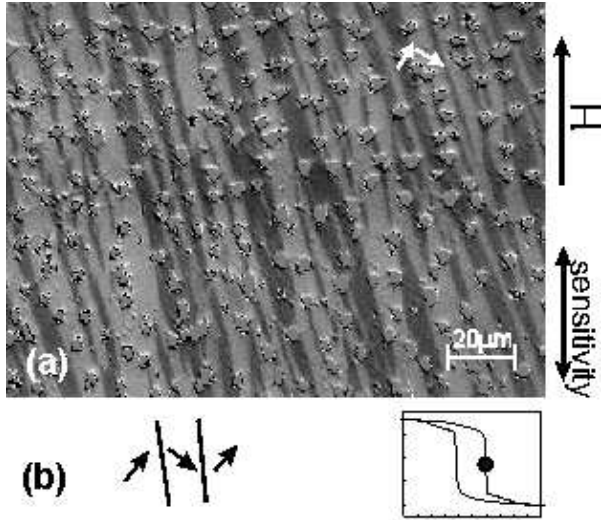}
\end{center}
\caption{%
\label{f:cr:outpeak}%
Kerr microscopy image of CCFA/Cr at H=\unit[21]{Oe} at a sample
orientation of $\alpha \approx 3^\circ$ ('out-of-peak'
orientation). The image contrast corresponds to the magnetization
component parallel to $H$. Magnetization directions are indicated
by small white arrows. (b) Sketch of the observed stripy domain
structure.}
\end{figure}

We now address the origin of the peaks in the coercivity of
CCFA/Cr. Figure~\ref{f:cr:outpeak}(a) shows a Kerr image of
CCFA/Cr at an angle of $\alpha \approx 3^\circ$ (i.e. in an
'out-of-peak' orientation), after the sample had been saturated in
a negative field and a field of \unit[21]{Oe} had subsequently
been applied. The image was recorded with sensitivity to the
magnetization component parallel to the direction of the applied
magnetic field. For the in-plane transverse magnetization
component we did not get any contrast, i.e. this magnetization
component is homogeneous. Hence, magnetization reversal solely
happens by the appearance of a stripy domain structure, as it is
sketched in Fig.~\ref{f:cr:outpeak}(b). The magnetization within
different domains points in different easy axis directions in such
a way that the stripy domains are separated by 90$^\circ$ domain
walls.

\begin{figure}
\begin{center}
\includegraphics[scale=1]{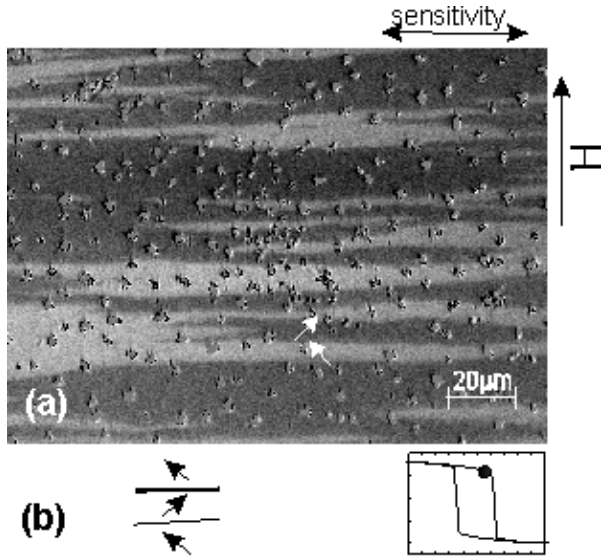}
\end{center}
\caption{%
\label{f:cr:atpeak}%
Kerr microscopy image of CCFA/Cr at H=\unit[18]{Oe} at a sample
orientation of $\alpha = 0^\circ$ ('peak' orientation). The image
contrast corresponds to the magnetization component transverse to
$H$. Magnetization directions are indicated by small white arrows.
(b) Sketch of the stripy domain structure.}
\end{figure}

\begin{figure}
\begin{center}
\includegraphics[scale=0.6]{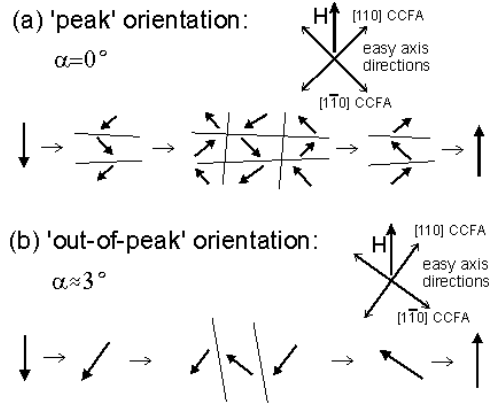}
\end{center}
\caption{%
\label{f:cr:sketch}%
Sketch of the magnetization reversal when (a) $H$ is parallel to
the $\langle$100$\rangle$ CCFA direction ('peak' orientation) and
(b) when $H$ is slightly deviated from the $\langle$100$\rangle$
CCFA direction ('out-of-peak' orientation).}
\end{figure}

Figure~\ref{f:cr:atpeak}(a) shows the reversal mechanism of
CCFA/Cr when $H$ is applied along the $\langle$100$\rangle$ CCFA
hard axes direction ($\alpha=0^\circ$). The sample had been
saturated in a negative field and then the image was taken at a
field value of \unit[18]{Oe}, i.e. before the jump in the
hysteresis loop appears. It should be noted that at this field
value no contrast was found for the magnetization component
parallel to $H$. Therefore, the domain configuration consist of
stripes [sketched in Fig.~\ref{f:cr:atpeak}(b)], which are again
separated by 90$^\circ$ domain walls, but now the stripe direction
is \emph{transverse} to $H$. The observed stripy domain structure
originates from a magnetic frustration effect, i.e. upon reduction
of the external field the magnetization wants to rotate into the
direction of an easy axis, but has to choose between two of these
axes which are energetically equivalent. Therefore the
magnetization splits into domains corresponding to these axes, as
sketched in Fig.~\ref{f:cr:atpeak}(b). If the external field is
now further reduced, a jump in the respective hysteresis loop
occurs, which is caused by the appearance of a 90$^\circ$ domain
structure, as it is sketched in Fig.~\ref{f:cr:sketch} (a).
Thereafter, the magnetization again assumes a stripy domain
configuration with a stripe direction transverse to $H$. The
reversal mechanisms in both the 'peak' ($\alpha = 0^\circ$) and
the 'out-of-peak' orientation ($\alpha \approx 3^\circ$) are
schematically illustrated in Fig.~\ref{f:cr:sketch}.

During hard axis magnetization reversal of CCFA/Fe, no stripy
domains transverse to the applied field $H$ were observed, i.e.
the only stripy domains observed were oriented parallel to $H$.
Therefore, this magnetization reversal process is similar to the
'out-of-peak' reversal of CCFA/Cr sketched in
Fig.~\ref{f:cr:sketch}(b).

\begin{figure}
\begin{center}
\includegraphics[scale=1]{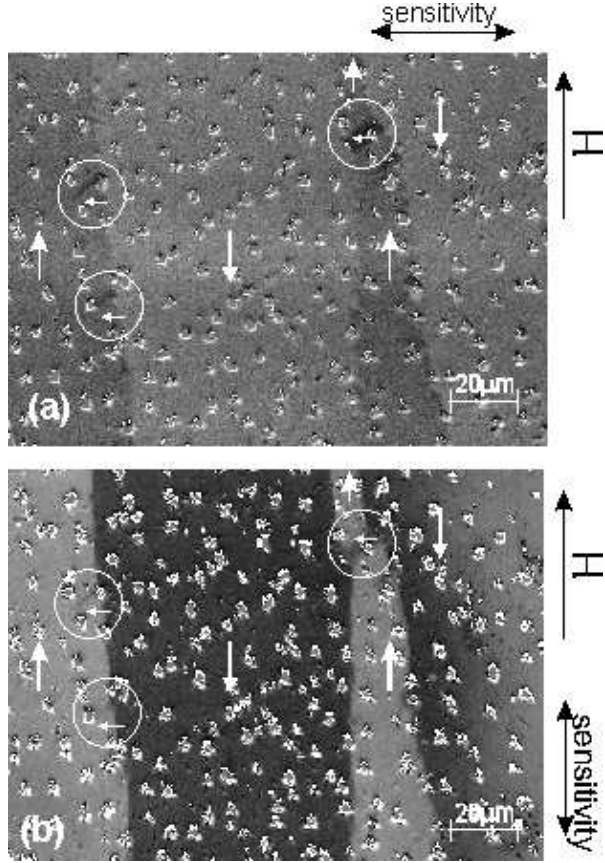}
\end{center}
\caption{%
\label{f:cr:easy}%
Kerr microscopy images of CCFA/Cr at H=\unit[29]{Oe} corresponding
to magnetization components (a) transverse and (b) parallel to $H$
applied along the [110] CCFA easy axis direction
($\alpha=45^\circ$). Magnetization directions are indicated by
small white arrows. Transversal spike domains are highlighted by
white circles.}
\end{figure}

Figure~\ref{f:cr:easy} displays snapshots of the magnetization
reversal process of CCFA/Cr in easy axis direction
($\alpha=45^\circ$). Panels (a) and (b) correspond to
magnetization components transverse or parallel to the applied
magnetic field, respectively. Magnetic domains are now mostly
separated by 180$^\circ$ domain walls. However, in this case
transverse spike domains containing a non-vanishing transverse
magnetization component appear (highlighted by white circles in
Fig.~\ref{f:cr:easy}). In the case of CCFA/Fe, such domains are
not observed.

\section{Summary}

The magnetic properties of Co$_2$Cr$_{0.6}$Fe$_{0.4}$Al (CCFA)
films deposited onto Fe and Cr buffer layers were studied. Both
films exhibit a fourfold magnetic anisotropy as well as a
90$^\circ$ domain structure with 90$^\circ$ domain walls in their
demagnetized state. Magnetization reversal occurs through the
formation of different stripy domain or 90$^\circ$ domain
structures. Furthermore, CCFA/Cr exhibits sharp peaks in $H_c$ if
$H$ is applied parallel to the $\langle$100$\rangle$ CCFA hard
axes directions. These peaks are related to the appearance of
peculiar domain structures during reversal originating from a
magnetic frustration effect. In the case of CCFA/Fe, such effects
are not observed.

\section{Acknowledgement}

The project was financially supported by the Research Unit 559
\emph{"New materials with high spin polarization"} funded by the
Deutsche Forschungsgemeinschaft, and by the Stiftung
Rheinland-Pfalz f\"ur Innovation. We would like to thank C. Hamann
for help with the Kerr microscopy measurements and T. Mewes for
stimulating discussions.

\newpage

\end{document}